\documentclass[journal=jacsat,manuscript=article]{achemso}
\usepackage[version=3]{mhchem} 

\usepackage{xcolor}
\usepackage[normalem]{ulem} 

\usepackage{multirow}
\usepackage{tablefootnote}
\usepackage{caption}
\usepackage{subcaption}
\usepackage{multicol}
\usepackage{array}

\usepackage{graphicx} 
\usepackage{float} 
\usepackage{listings} 
\usepackage{xcolor}
\usepackage{tikz}
\title{Quantitative Analysis of \ce{MoS2} Thin Film Micrographs with Machine Learning}
\author{Isaiah A. Moses}
\affiliation[MRI]{Materials Research Institute, The Pennsylvania State University, University Park, PA 16802}
\author{Wesley F. Reinhart}
\email{reinhart@psu.edu}
\affiliation[MTSE]{Department of Materials Science and Engineering, The Pennsylvania State University, University Park, PA 16802}
\alsoaffiliation[ICDS]{Institute for Computational and Data Sciences, The Pennsylvania State University,
University Park, PA 16802}

\begin{document}

\maketitle

\begin{abstract}
Isolating the features associated with different materials growth conditions is important to facilitate the tuning of these conditions for effective materials growth and characterization. 
This study presents machine learning models for classifying atomic force microscopy (AFM) images of thin film \ce{MoS2} based on their growth temperatures.
By employing nine different algorithms and leveraging transfer learning through a pretrained ResNet model, we identify an effective approach for accurately discerning the characteristics related to growth temperature within the AFM micrographs.
Robust models with test accuracies of up to 70\% were obtained, with the best performing algorithm being an end-to-end ResNet fine-tuned on our image domain.
Class activation maps and occlusion attribution reveal that crystal quality and domain boundaries play crucial roles in classification, with models exhibiting the ability to identify latent features that humans could potentially miss.
Overall, the models demonstrated high accuracy in identifying thin films grown at different temperatures despite limited and imbalanced training data as well as variation in growth parameters besides temperature, showing that our models and training protocols are suitable for this and similar predictive tasks for accelerated 2D materials characterization.
\end{abstract}

\section{Introduction}

Material properties are significantly influenced by conditions experienced during synthesis.\cite{13zavabeti2020two, 14lei2022graphene, 1eichfeld2015highly,5zhang2016influence, 15zhang2018diffusion}
A systematic way of isolating the properties associated with different conditions is essential to enable the growth of materials with predefined properties on demand.
We particularly seek approaches that eliminate intuition-based experimentation with different process variables, replacing them with data-driven approaches that are more efficient with time, effort, and other resources.

Several studies on thin film \ce{MoS2} have revealed a number of growth parameters that determine the morphological features and properties of the grown materials.
Instances include the evolution of the morphology of monolayer \ce{MoS2} crystals grown by chemical vapor deposition (CVD).\cite{wang2014shape}
Domain shape variation from the triangular to hexagonal geometries has been shown to depend on the Mo:S ratio of the precursors.\cite{wang2014shape}
Similarly, a \ce{MoS2} domain shapes of mainly round, nearly round and hexagonal, truncated triangles, and triangles are observed at the temperatures of the \ce{MoO3} precursor of 760$^\circ$C, 750$^\circ$C, 730$^\circ$C, and 710$^\circ$C, respectively.\cite{xie2016high}

The density and size of the domain have also been shown to decrease with temperature,\cite{xie2016high, suleman2022nacl} with a random orientation of the \ce{MoS2} domain associated with the growth temperature below 850$^\circ$C\cite{2li2021epitaxial} or at a much higher temperature.\cite{3xiang2020monolayer}
In the former, the authors linked the phenomenon to the inability to achieve a thermodynamically stable state at the lower temperature, and in the latter, the inferred culprit is the step edges and step edge meanderings of sapphire substrate surface.

The grain size and crystal coverage of the the \ce{MoS2} have also been shown to be tunable with the growth time.\cite{xie2016high}
The authors showed that the grain size increased when the growth time was increased from 20 minutes to 30 minutes.
With the materials grown for 45 minutes, the grains merged to form a continuous \ce{MoS2}.\cite{xie2016high}
Similarly, an increased in growth temperature\cite{suleman2022nacl} and \ce{O2} flow rate\cite{yang2022oxide} were shown to result in larger thin film crystal coverage.

In designing high throughput on-demand materials, deployment of data-based screening approaches have become more critical.\cite{23han2020machine, 24gu2022perovskite, moses2021machine, moses2022accelerating, 25yan2021accelerated, 26lu2022fly}
Data-driven approaches are being explored for materials characterization\cite{yang2020automated, han2020deep, jung2023automatic, si20232d, saito2019deep} and serve to provide greater clarity when searching the synthesis condition space compared to intuition-based experimentation.\cite{21tang2020machine, 22beckham2022machine, lu2022machine, frey2019prediction, ryu2022understanding}
With the use of the existing data consisting of the conditions and the corresponding materials properties, models that predict what conditions are necessary for a given properties can be developed.
As observed, a number of these conditions play similar and intertwined roles in the materials properties.
For instance, the time, temperature, and \ce{O2} flow rate determine the \ce{MoS2} thin film crystal coverage.\cite{xie2016high, suleman2022nacl, yang2022oxide}
It will be interesting to use machine learning to isolate the distinct latent features associated with the different growth parameters.
Additionally, identifying distinct latent features for these different growth parameters would result in the capability to classify material samples based on their growth conditions. 

The Lifetime Sample Tracking (LiST) is a database hosted by the Penn State 2D Crystal Consortium (2DCC) facility, consisting of experimentally grown thin film transition metal chalcogenides materials, among others.
Among the characterization methods used in the 2DCC and stored in LiST is Atomic Force Microscopy (AFM).
AFM micrographs of \ce{MoS2} thin films and their corresponding synthesis conditions are a set of data among other categories in LiST.\cite{3xiang2020monolayer, 27schranghamer2023ultrascaled, 28trainor2022epitaxial}
To accelerate the synthesis of \ce{MoS2} with the desired properties, we deploy different machine learning (ML) models to classify AFM images of the material based on their growth temperature. The ultimate goal of the machine learning models is for the inverse design of materials, where the materials properties are tuned using the growth parameters. In essence, being able to predict the growth conditions from the morphology will enable the ability to determine the best growth conditions to achieve a hypothetical film morphology. This should accelerate the design and tuning of materials synthesis in the future.
Despite the limited data available for the training, up to 71\% test accuracy was obtained on the image classification.
Most importantly, this study presents a simple approach that could help isolate underlying morphological features associated with different growth conditions for a broad range of materials, paving the way for rapid and cost-effective materials development.

\section{Methods}

\subsection{Data Preparation}

Raw \texttt{spm} files of \ce{MoS2} were retrieved from LiST.\cite{Moses_Reinhart}
These 262 AFM height maps were processed into greyscale images and either resized or randomly cropped to the common size of $224 \times 224$, depending on the augmentation method adopted, as discussed below.
Training computer vision models on such a small dataset requires transfer learning, a common approach that utilizes CNN models pretrained on one image domain to extract features from a new image domain.\cite{29kitahara2018microstructure, 30gong2020deep, 31cohn2021unsupervised}
Many popular pretrained CNNs, such as the VGG,\cite{VGGsimonyan2014very} ResNet,\cite{RESNEThe2016deep} and Inception model\cite{INSPECTIONszegedy2015going, INSPECTION2szegedy2016rethinking} architectures were trained on the ImageNet dataset.\cite{IMAGENETdeng2009imagenet}
ImageNet contains millions of color images of natural objects from thousands of categories.
Using the size of the model architecture as the main basis for our choice, because of the small data volume in our characterization problem, the ResNet18 architecture pre-trained on ImageNet is used for transfer learning.

\begin{figure}
    \centering
    \includegraphics{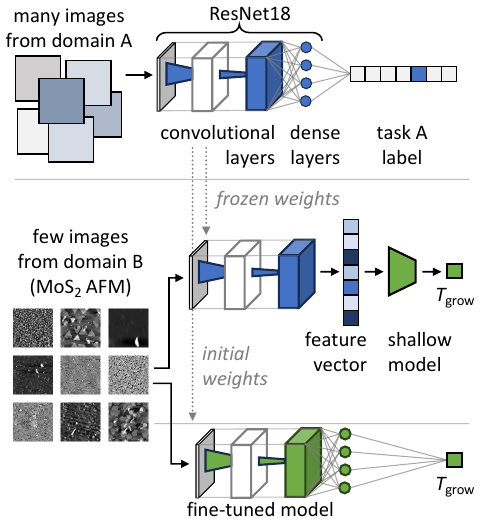}
    \caption{
    An overview of the transfer learning approach.
    (top) A ResNet CNN model is trained on a different image domain with a large number of images.
    The task may be unrelated to the present task -- all that matters is that convolutional filters are learned that can extract information (e.g., texture, color, shapes) from the images.
    (middle) The filters from the pretrained model can be used directly to extract relevant image features, which are interpreted in a supervised manner by a shallow model to predict a new label, such as the growth temperature.
    (bottom) Alternatively, the filters from the pretrained model can be fine-tuned on the new image domain to better capture relevant information for the task at hand.
    }
    \label{fig:schematic}
\end{figure}

However, our data distribution is very different than the ImageNet data.
To evaluate the effect of the pretraining domain, we consider pretraining on micrographs contained in the MicroNet dataset\cite{MICRONETstuckner2022microstructure}, which should be more similar to our image domain.
The MicroNet dataset has been shown to give better performance on micrographs, indicating that the proximity of the two image domains should enhance the model performance.\cite{MICRONETstuckner2022microstructure}
We have therefore additionally used ResNet18 pretrained on the MicroNet dataset.
This will enable us to compare how the same model architecture pretrained on different datasets perform on our characterization task.
Features were extracted from the pretrained models for our shallow ML models.
The pretrained convolutional models were also fine-tuned for the CNN model in our study (Figure \ref{fig:schematic}).

\subsection{Data Augmentation}

\begin{figure}
 \centering
     \includegraphics[width=\textwidth]{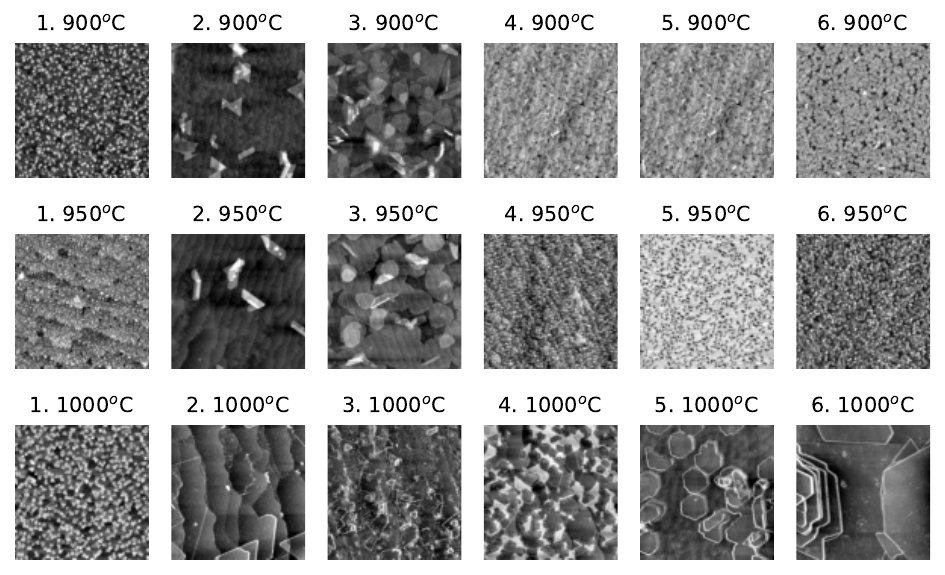}
     \caption{Sample images from \ce{MoS2} grown at 900, 950, and 1000$^\circ$C.}
     \label{sample_AFM}
\end{figure}

The dataset consists of 262 instances of AFM height maps across 3 growth temperatures (Figure \ref{sample_AFM}). 
In addition to the limited data, there is a significant imbalance among the different classes with the 900$^\circ$C, 950$^\circ$C, and 1000$^\circ$C making up 11\%, 50\%, and 39\% respectively (Table \ref{augtab}). 

\begin{table}[ht]
	\centering
	\caption{Data augmentation policies and the corresponding data sets for the different classes, 900$^\circ$C, 950$^\circ$C, and 1000$^\circ$C.
    In \textit{Aug1}, multiple random cropping of image size $224 \times 224$ is used to obtain balanced instances among the different classes, \textit{Aug2} is augmentation policy learned on ImageNet\cite{19cubuk2019autoaugment}, and in \textit{Aug3} weighted random sampler and oversampling are used to correct the imbalance in train set for CNN and other models, respectively.
    \textit{Aug4} is without biased augmentation.
    In CNN models, random rotations between 0 to 180$^\circ$, horizontal and vertical flipping at 50\% probability were additionally used on the train and validation set on the fly.}
	\label{augtab}
		\begin{tabular}{l| c   c    c  |  c     c     c   |  c     c   c|   c} \hline
    &  \multicolumn{3}{c|}{\bf{900$^\circ$C}} & \multicolumn{3}{c|}{\bf{950$^\circ$C}} &\multicolumn{3}{c|}{\bf{1000$^\circ$C}} & \bf{Total}\\ 
    &   train&  validation  & test &  train&  validation  & test &  train&  validation  & test &   \\ \hline
    \it{Aug1}    &  207 &  27  &    3  &   212 &   22  &   13  &   215 &   24  &   11  &   726 \\
    \it{Aug2}    &  207 &   27  &   3  &   208 &   24  &   13  &   210 &   23  &   11  &   734 \\
    \it{Aug3}    &   105 &   3   &   3   &   105 &   12  &   13  &   105 &   9  &   11  &   342 \\
    \it{Aug4}    &   23  &   3   &   3   &   105 &   12  &   13  &   83  &   9  &   11  &   262 \\ \hline
    \end{tabular}
\end{table}

The effect of limited and imbalanced data on the model performance can be partially mitigated with data augmentation approaches.
Different data augmentation policies were therefore deployed to determine which method works best for our small, imbalanced dataset.
The first was to randomly crop a common size of $224 \times 224$ from each of the original images.
Multiple croppings were carried out, depending on the class of the image, in order to obtain a balanced representation of the different classes.
This augmentation policy is termed \textit{Aug1} (Table \ref{augtab}).
Another augmentation policy examined is that developed by Cubuk, et al,\cite{19cubuk2019autoaugment} which we referred to as \textit{Aug2} hereafter.
The authors used a search algorithm to find the best policy, which is a combination of many sub-policies consisting of functions such as the translation, rotation, or shearing, and the probabilities and magnitudes with which the functions are applied, that give the best validation accuracy on a target dataset.
Interestingly, they observed that the learned policy in a given dataset is transferable to another.
We therefore examined how transferable the policy learned on ImageNet is to our present data domain.
The third augmentation method used is a weighted random sampler or oversampling to correct the imbalance in the training set (\textit{Aug3}).
For \textit{Aug4}, there is no biased augmentation applied to the data and only in CNN models do we have random rotations between 0 to 180$^\circ$, horizontal and vertical flipping at 50\% probability applied to the train and validation set on the fly.

\subsection{Machine learning}

A 10-fold cross-validation training scheme was used to train and evaluate the models, with 10 different models trained, one for each train-validation data splitting. 10 \% of the data was held out for testing while, 90 \% was randomly split into 10 equal folds. A unique fold was used for the validation (to determine the performance for hyperparameter tuning using grid search) in each of the 10 models while the remaining 9 folds were used for training model parameters.
The hyperparameters of the model with the best performance from the cross-validation procedure were selected for the production model. The 10 different training sets were then fitted independently into the production model and a held-out test set (not involved in the cross-validation procedure) was then used to evaluate the model performance in general.

Nine different ML models were considered:
support vector classifier (SVC)\cite{SVMchang2011libsvm, SVMplatt1999probabilistic},
kernel ridge classifier (KRC)\cite{KRmurphy2012machine},
radius neighbors classifier (RNN)\cite{NNgoldberger2004neighbourhood},
Gaussian process classifier (GPC)\cite{GPrasmussen2006gaussian},
k-nearest-neighbors classifier (KNN)\cite{NNgoldberger2004neighbourhood},
decision tree classifier (DTC)\cite{DRandRFcutler2012random},
gradient boost classifier (GBC)\cite{GBfriedman2001greedy},
multilayer perceptron (MLP)\cite{MLPhornik1989multilayer},
and convolutional neural network (CNN)\cite{CNNlecun1989handwritten, CNNlecun1998gradient}.
The shallow models were developed using the \texttt{scikit-learn} library version 1.2.2\cite{SCIKITLEARNpedregosa2011scikit} and the MLP and CNN were implemented in \texttt{pytorch}\cite{PYTORCHpaszke2019pytorch}. The optimized hyperparameters for the models are shown in the Supporting Information. The MLP model consists of 2 hidden layers, with each followed by a ReLU activation function. Additionally, we placed a drop out layer just before the output layer. For the CNN (fine-tuned pretraind ResNet model), the classifier outputs 3 classes for classification, but is replaced with a 100 nodes fully connected layer and an output layer for the regression models.

Using AFM images of 2D \ce{MoS2} grown with MOCVD, we developed models to predict the growth temperature (one of 900$^\circ$C, 950$^\circ$C, or 1000$^\circ$C).
We considered framing the task in several different ways to evaluate the efficacy of each: nominal classification, ordinal classification, and regression.
Here nominal classification means the three growth temperatures were considered as distinct classes with no ordering.
Unless otherwise specified, results are for nominal classifiers.

For ordinal classification, we implement NNRank\cite{20cheng2008neural} to account for ordering within the classes; the targets 900, 950, and 1000$^\circ$C are transformed into the vectors $[1, 0, 0]$, $[1, 1, 0]$, and $[1, 1, 1]$, respectively.
At inference time, a threshold of $>0.5$ is applied to the prediction and the values are counted from left to right, which provides the class label.
Note that this scheme is only applied to the NN models (MLP and CNN).
Finally, we perform regression by simply using the growth temperatures as continuous labels and evaluating the MSE.
The class labels are obtained by binning the predicted growth temperature (e.g., $925 - 975^\circ$C belongs to the $950^\circ$C class).

\section{Results and Discussion}

\subsection{Depth of Image Features}

\begin{table}[ht]
	\centering
	\caption{
Validation accuracy (in \%) based on the features extracted from the different layers of the pretrained model (ResNet18 pretrained on ImageNet).
 Channels is the total size of raw feature vectors extracted from each block of the ResNet.
 PCA was applied to these channels, and then cumulative explained variance (CEV) of the components from PCA was used to determine the size of the input features for the listed shallow models.
 Separately, the dense layers of the pretrained model were replaced with fewer neurons and fine-tuned (last column).
 }
	\label{layer}
		\begin{tabular}{l|c    c |   c   c   |  c     c   |  c     c |  c  c } \hline
& \multicolumn{2}{c|}{\bf{Block 2}}  &\multicolumn{2}{c|}{\bf{Block 3}}  & \multicolumn{2}{c|}{\bf{Block 4}}   & \multicolumn{2}{c|}{\bf{Pooling}}& \bf{Fine-Tuned}  \\
Channels & \multicolumn{2}{c|}{100352} & \multicolumn{2}{c|}{50176} & \multicolumn{2}{c|}{25088} & \multicolumn{2}{c|}{512} & 100 \\ \hline
CEV & 85\%  &  99\%  & 85\%  &  99\%  &  85\%  &  99\%  &  85\% &  99\%  & -  \\
Features & 190 & 235 & 156 & 235 & 94 & 219 & 28 & 142 & 100 \\
\hline
SVC& 66$\pm$6   & 59$\pm$7 &64$\pm$5   & 62$\pm$8    &   77$\pm$6  &    58$\pm$5  &   78$\pm$5  &  71$\pm$6   &    80$\pm$7   \\
KRC & 45$\pm$5 & 57$\pm$4 & 48$\pm$11   & 55$\pm$5    &   58$\pm$11  &    52$\pm$5  &   57$\pm$7  &58$\pm$12   &   71$\pm$7  \\
RNN & 21$\pm$4 & 15$\pm$2 & 35$\pm$11   & 15$\pm$3    &   42$\pm$9  &    20$\pm$5  &   57$\pm$7  & 39$\pm$7   &   70$\pm$9 \\ \hline
	   \end{tabular}
\end{table}

Given the poor performance observed from the randomly initially weights of the CNN models (Supporting Information), we deployed transfer learning for the task. We first determined the best location in the pretrained model from which to extract image features for our models.
Different portions (``blocks'') of the ResNet were considered, providing filters with different levels of abstraction.
Due to the large number of channels in the pretrained model (see Table~\ref{layer}), Principal Component Analysis (PCA) was applied to reduce the dimension of input features to the shallow models, ideally reducing overfitting and thus improving predictive performance.\cite{PCAdoi:10.1080/14786440109462720, PCAjolliffe2016principal, moses2021machine}
Cumulative explained variance thresholds of 85\% and 99\% were used to determine the number of features to keep for inference.
We found that within a block, using fewer features gave better performance in 9 of 12 cases despite lower explained variance, likely because we had few training data compared to the size of the feature vectors.
Depending on the model architecture and number of features used, minimal or significant deviations in model performance could be obtained from any of the ResNet blocks (e.g., 66\%, 64\%, 77\%, and 78\% accuracy from subsequent blocks, with typical standard deviation $\pm 6\%$).

Separately, the dense layers of the pretrained model was replaced with new ones with fewer neurons and then fine-tuned on our training data. The model parameters are the same as the CNN classifier described in the previous section. The model fine-tuned on the ImageNet and the MicroNet gave a train accuracy of 88 and 76 \%, respectively, and a validation accuracy of and 73  and 70 \%, respectively.
Finally, 100 features were extracted from the first dense layer. Note that we have compared the performance of this fine-tuned dense layer against those extracted from the pretrained blocks. This was an intentional choice to evaluate the degree to which fine-tuning was needed to achieve good performance in this task.

The performance of the selected classifiers on the different features shows that the features extracted from the fine-tuned dense layer gives the best performance overall, with 80\%, 71\%, and 70\% accuracy using SVC, KRC, and RNN, respectively.
Training the dense layer on a pretrained convolutional backbone might therefore be a better approach for extracting a low-dimensional image feature vector compared to PCA.
These tuned features are therefore used in all of the following analysis.

\subsection{Data Augmentation}

\begin{figure}
     \centering
         \includegraphics[width=\textwidth]{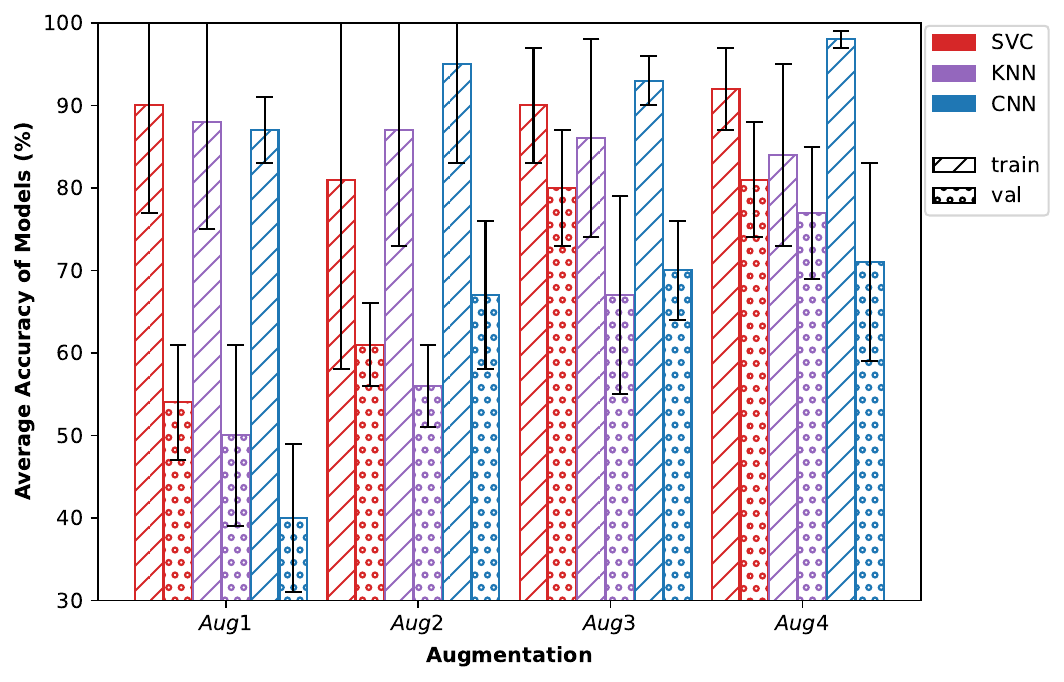}
         \caption{Accuracy obtained from different augmentation policies across three different model types.
         Bars report averages over 10 folds, while error bars indicate standard deviation.
         Some models were trained with increased data size to have a balanced classes using different augmentation approaches, as indicated in Table~\ref{augtab}.}
         \label{augfig}    
\end{figure}

We then evaluated the effect of different data augmentation policies using the SVC, KNN, and CNN models (Table~\ref{augtab} and Figure~\ref{augfig}). In addition to the accuracies of the models, F1 score was used to evaluate the different augmentation policies. This is to ensure that the data imbalance is accounted for in comparing their performances. It is observed that both the accuracy and F1 score gave similar performance trend (Figure~\ref{augfig} and Figure S2).
Significantly worse performances are obtained with \textit{Aug1} and \textit{Aug2}, especially in the shallow models, compared to \textit{Aug3} and \textit{Aug4}.
Meanwhile, the performance observed between \textit{Aug3} and \textit{Aug4} is statistically indistinguishable.

The poor performance observed in the \textit{Aug1} and \textit{Aug2} might be related to the properties of the images learned by the models.
While in the case of the natural images, activation of different classes are typically associated with unique features of the classes,\cite{CAM8004993, CAM9206626, CAM9462463} the class activation in the models for the different synthesis conditions will be more likely due to differences in magnitude of the same feature, such as the domain size and thickness.\cite{1eichfeld2015highly, 5zhang2016influence}
These relevant features of the AFM images may be disrupted by shearing, zooming, and resizing associated with \textit{Aug1}, and the features location in the image might be omitted due to the cropping in \textit{Aug2}. 

Although \textit{Aug3} and \textit{Aug4} present about the same accuracy, \textit{Aug3} has the desirable property of oversampling less represented classes.
This should help mitigate systematic error related to class imbalance, a feature which is typical of distributions in materials synthesis, especially when exploring different growth conditions (e.g., poorly performing conditions will probably be undersampled).
Therefore, the \textit{Aug3} augmentation policy is selected for the rest of this study.

\subsection{Pretraining Domain}

\begin{table}[ht]
	\centering
	\caption{Validation accuracy (in \%) over 10 folds obtained for the feature extraction (shallow and MLP models) or end-to-end learning (CNN) with ResNet18 pretrained on ImageNet and MicroNet.
    Values are reported as mean $\pm$ standard deviation.
    Difference is the fractional change in the average score between MicroNet and ImageNet.
    Best model performance in each row is shown in bold.
    }
	\label{features}
		\begin{tabular}{l|c    c    c   c     c     c     c     c   c   } 
Models    &SVC      & KRC       &  RNN   & GPC   & KNN   & DTC   & GBC   & MLP   &  CNN \\\hline
MicroNet  & \textbf{73$\pm$6} & 65$\pm$10 & 63$\pm$9  & 52$\pm$12    &   59$\pm$10  &   71$\pm$9  &  71$\pm$9  &  65$\pm$8   &    63$\pm$8  \\
ImageNet    &80$\pm$7 & 71$\pm$7 & 70$\pm$9   & 59$\pm$10    &  67$\pm$12   &   78$\pm$4  &   78$\pm$11  &   \textbf{86$\pm$6}  &    70$\pm$6   \\\hline
Difference & +10\% & +9\% & +11\% & +13\%  & +14\% & +10\%  &  +10\%  &  +32\% &  +11\% \\ \hline
	   \end{tabular}
\end{table}

The previous two sections on the feature extraction and the data augmentation are initial verifications. Therefore, only 3 machine learning models were explored. We next seek to quantify how transfer learning from the ResNet18 model pretrained on the ImageNet data domain compares with the same model architecture pretrained on the seemingly more relevant MicroNet data domain.
We therefore compared the performance of each pretrained model on the same nominal classification task across a wide range of predictive model types.
In these experiments, we used the fine-tuned features from Table~\ref{layer} in all cases except CNN, which was simply fine-tuned in an end-to-end manner using the original ResNet18 architecture (i.e., with a three-way classification layer attached to the end in place of the original classification layer).
Based on the results shown in Table~\ref{features}, the ImageNet model gives conclusively better performance than MicroNet, with at least 9\% improvement and up to 32\% improvement in the case of MLP (compared to a typical uncertainty of about 6\%).

While standard deviations for individual observations are high, the fact that none of the nine model types shows a negative difference is compelling, especially because MicroNet was trained on greyscale micrographs of materials while ImageNet was trained on color images of macroscale objects.
Previous work has suggested that ImageNet relies more heavily on texture rather than shape\cite{geirhos2018imagenet}, while MicroNet has been primarily tested for segmentation tasks.
We speculate that this focus on texture gives ImageNet filters that can be used for identifying distinguishing textures in the AFM height maps.
The results presented here suggest that ImageNet may be surprisingly well suited for out-of-domain materials characterization data whose information content is primarily texture.
All following results are based on transfer learning from the ImageNet pretraining since its features are strictly superior to MicroNet.

\subsection{Model Performance}

\begin{figure}
     \centering
     \includegraphics[width=\textwidth]{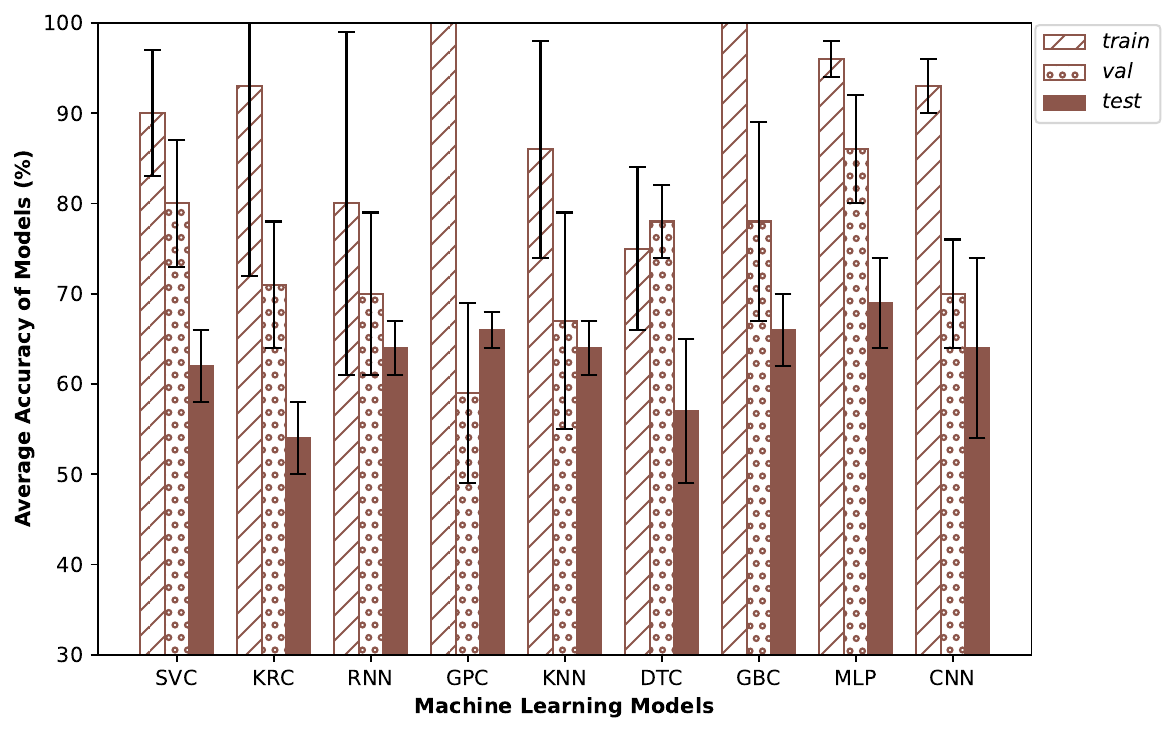}
         \caption{The average train, validation (\textit{val}), and test accuracy over 10 models for the different algorithms.
         The train-validation data was randomly split into 10 equal folds. A unique fold was used for the validation in each of the 10 models while the remaining 9 folds were used for training model parameters.
         Hyperparameters were tuned to obtain a trained model for each of the 10 splits.
         The trained models were tested with the test set.}
         \label{val}
\end{figure}

We next investigate the performance of different algorithms in greater detail.
As before, we rely on the features extracted from the fine-tuning procedure above, with additional shallow models trained on these static feature vectors of each image.
The CNN model is the one exception to this, as it uses the original ResNet18 architecture and is fine-tuned on this task without modification to feature size.
The classification accuracy across 10 different model instances of each type is shown in Figure~\ref{val}.
Overfitting is observed across all model types, with training performance over 90\% being typical, while validation typically only reaches around 60-85\%.
The greatest overfitting, in terms of the gap between train and validation performance, is seen in KRC and GPC, while SVC, DTC, and MLP exhibit the least.
The best performing models in terms of validation performance is the MLP, with SVC coming in second but exhibiting training and validation scores one standard deviation below the MLP.

To understand how well the models can generalize to classifying images outside of the training data, we additionally examine their performance on a held-out test set (i.e., not used for training or hyperparameter selection).
In this regard, MLP again showed the highest accuracy, with GBC and GPC appearing within one standard deviation.
It is reassuring to see that MLP gave the highest scores in both validaiton and testing, inspiring confidence in its performance overall.

\begin{figure}
     \centering
     \includegraphics[width=\textwidth]{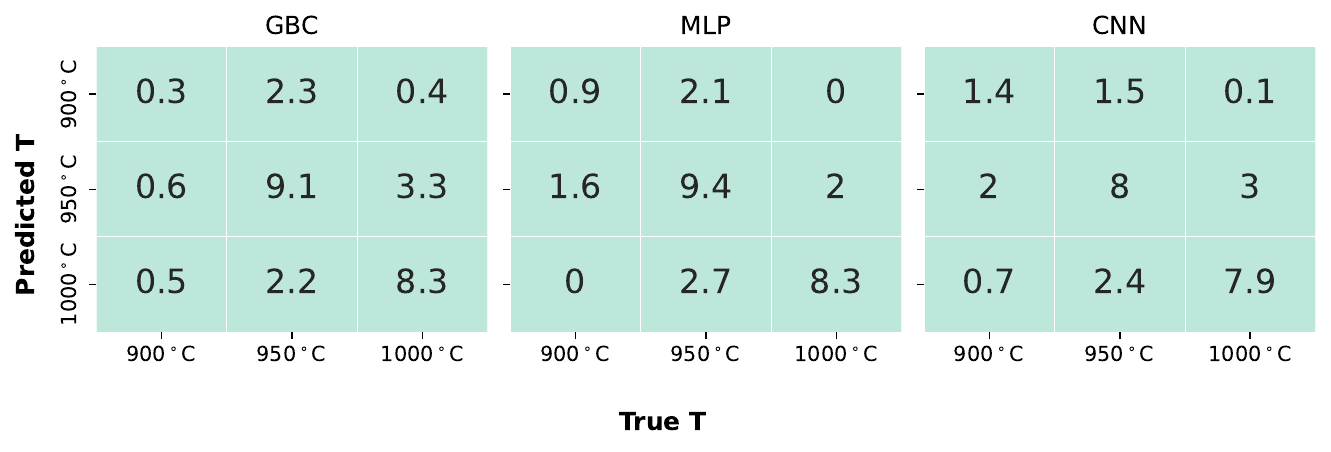}
         \caption{The average confusion matrix for the test set predictions of production models trained on the 10 folds train data.
         Values indicate the number of samples in each bin.
         This is based on nominal classification.
         }
         \label{testcm}
\end{figure}

To understand the model performance on the different growth temperatures in greater detail, and particularly to check if the underrepresented classes have comparable accuracy, average confusion matrices of the held-out test set on 10 models are reported in Figure~\ref{testcm}.
To focus the discussion, only the highly performant GBC and MLP models and the end-to-end CNN are examined in this regard.
It is notable that the performance within each class does not vary substantially between different model types, as the overall accuracy are similar.
For instance, the GBC, MLP, and CNN predict about the same number of samples grown at 950$^\circ$C and 1000$^\circ$C correctly (about 70\% and 75\% respectively).
The samples grown at 900$^\circ$C are found to have the lowest in-class accuracy.
This seems to be partially an artifact of under-representation in the test set; as shown in Table~\ref{augtab}, classes are significantly imbalanced in the data, with the 900$^\circ$C classes having the least number of samples.

There is also some consistency among the models in misclassifying the 900$^\circ$C as 950$^\circ$C and not as 1000$^\circ$C.
Similarly, 1000$^\circ$C is rarely misclassified as 900$^\circ$C.
On the contrary, 950$^\circ$C is about equally likely to be misclassified as 900$^\circ$C as it is as 1000$^\circ$C by the MLP and CNN.
This seems to suggest that the proximity of the growth temperature, which is expected to be reflected in the image features, makes it more likely for the model to group them together.
Recall that this is for nominal classification, so this proximity is not reflected in the loss function.
This could imply a fundamental bias in the data where the image feature learned by the models for a given temperature are more similar to that for the adjacent temperatures.

\begin{figure}
     \centering
     \includegraphics[width=\textwidth]{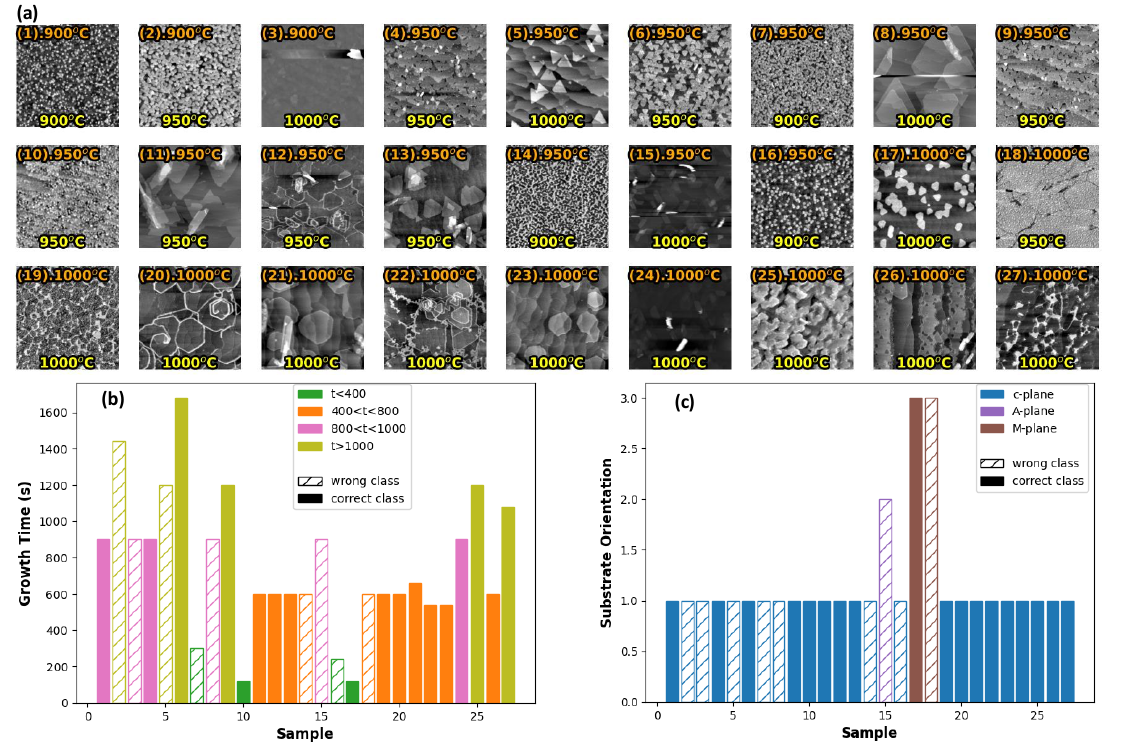}
         \caption{Samples grown at 900$^\circ$C (1-3), 950$^\circ$C (4-16), and 1000$^\circ$C (17-27) in the test set. The predicted class by end-to-end CNN is shown at the bottom (yellow) for each image.
         (b) and (c) are the samples with their growth time and substrate orientation, respectively.
         The AFM \# in (a) corresponds to the sample \# in (b) and (c).}
         \label{test_data}
\end{figure}

To further understand the classification fidelity of our models, we examine images that are correctly and incorrectly classified by the CNN in Figure~\ref{test_data}.
Visual inspection suggests significantly different image features among the same growth temperature, demonstrating how difficult this classification task is.
Some images grown at 950$^\circ$C show larger crystal domains typically associated with 1000$^\circ$C.
Conversely, some images grown at 1000$^\circ$C show poor crystal formation and very small domain sizes exhibited mostly by the 900$^\circ$C growth temperature.
Therefore, these wrongly classified images may be exceptional among the target class and would likely confuse even a human expert.
However, they offer some preliminary insight into which features the classifier attributes to each growth temperature.

More fundamentally, other growth variables are not entirely fixed across the samples.
For instance, the growth time varies significantly among the different samples (Figure \ref{test_data}(b)).
While the least growth time in the test set is as low as about 100~s, some samples are grown at much longer time, with up to 1650~s.
Also, while most of the samples are grown on c-plane sapphire substrate, we also have some that are grown on A- and M-plane sapphire (Figure \ref{test_data}(c)).
These inconsistent growth parameters might have accounted for the significant differences observed among the samples grown at the same temperature and might have also resulted in some classification errors (e.g images 2, 5, 15, and 18).
However, we do not observe any obvious trend in these growth parameters that leads to consistent misclassification, once again demonstrating how challenging this classification task is.

\subsection{Ordinality}

The preceding results were all based on nominal classification, without any notion of ordering.
However, the classes consisting of the growth temperatures would appear to be ordered due to their continuous nature (i.e., ranging from 900 to 1000$^\circ$C).
We therefore further quantify the effect of ordinal treatment of the class labels on model accuracy.
In accounting for ordinality in shallow (i.e., non-NN-based) models, we adopted a simple approach based on training a regressor and then binning the results into classes.
For the NN-based models, we further implemented the NNrank ordinal classification scheme.
The results of this study are given in Table~\ref{ordi}.

\begin{table}[ht]
	\centering
	\caption{Performance of nominal and ordinal treatment of class labels, expressed as accuracy on held-out test data in \% for classification and $^\circ$C for regression.
    Best model performance in each row is shown in bold.
    }
	\label{ordi}
		\begin{tabular}{l c    c    c   c     c     c     c     c   c } 
  \hline
Models  &SVM   & KR   & RNN   & GP   & KNN   & DT   & GB   & MLP   &  CNN \\\hline
Classification (\%) &62$\pm$4 & 54$\pm$4 & 64$\pm$3   & 66$\pm$2    &   64$\pm$3   &   57$\pm$8  &  66$\pm$4  &   \textbf{69$\pm$5}  &    64$\pm$10 \\
NNRank (\%) & - & - & - & - & - & - & - &  \textbf{71$\pm$4}   &    68$\pm$3  \\
Regression (\%) & 50$\pm$8 & 60$\pm$5 & 64$\pm$4   & 48$\pm$0    &   58$\pm$6  &    54$\pm$6  &   \textbf{64$\pm$7}  &  42$\pm$8   &    61$\pm$7  \\
\hline
RMSE ($^\circ$C) & 31$\pm$2 & \textbf{26$\pm$1} & - & 36$\pm$9 & 32$\pm$3 & 38$\pm$5 & 28$\pm$3 & 62$\pm$8 & 34$\pm$4 \\
\hline
	   \end{tabular}
\end{table}

While results vary for each model type, some general trends emerge.
Accounting for ordinality in model training leads to improvement in the test accuracy in only one of the shallow models (KR), but matches or degrades the performance for all others.
Most of these are statistically indistinguishable, with only SVM, GP, and MLP exhibiting significant decreases.
Overall, nominal classification gave superior performance over regression, with the top performing shallow models GP and GB giving 66\% accuracy.

For the NN models, MLP outperformed CNN overall, with statistically indistinguishable accuracy using nominal classification and ordinal classification.
While the end-to-end CNN performed significantly better than the MLP on the regression task, the performance on regression was the worst of the three schemes for each model, making it somewhat irrelevant.
Somewhat counterintuitively, slightly higher accuracy could be obtained by binning the output of the GB regressor (64\%) which had a higher RMSE compared to the KR regressor ($28 \pm 3^\circ$C versus $26 \pm 1^\circ$C).
This suggests that least-squares regression may be placing too much weight on outliers, which are less influential in the case of ordinal classification.
It is even possible that the growth temperatures are not really ordinal after all, perhaps with 950$^\circ$C representing a value close to optimal while 900$^\circ$C and 1000$^\circ$C could be a similar distance away from optimal.

\begin{figure}
     \centering
     \includegraphics[width=\textwidth]{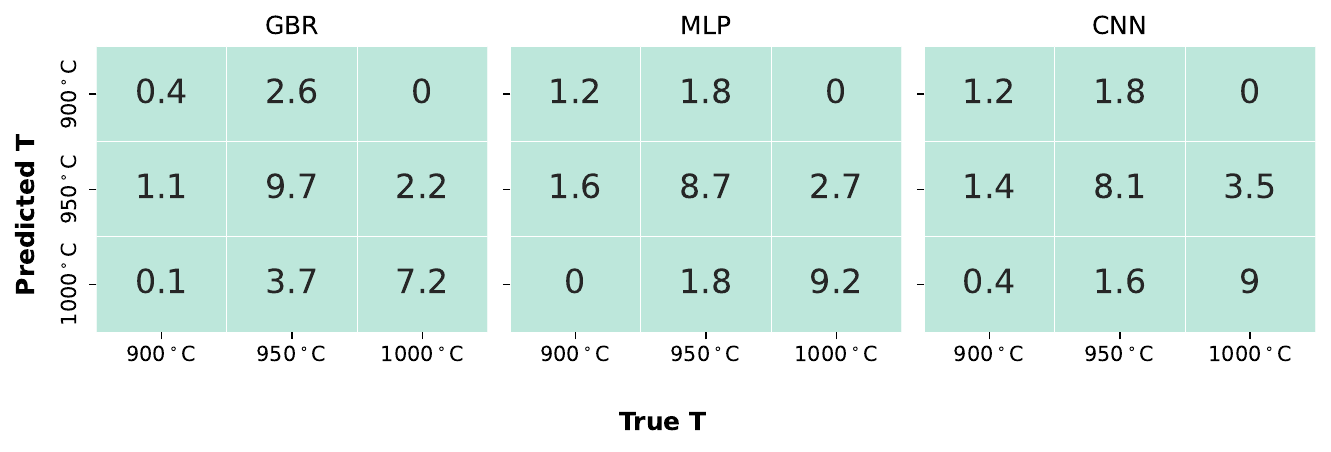}
         \caption{The average confusion matrix of the 3 classes of temperature (900, 950, and 1000$^\circ$C) on the test set for the 9 different model architectures.
         This is based on ordinal classification with regression used for the GBR and NNrank for the MLP and CNN.}
         \label{testcm_ordi}
\end{figure}

The best-performing model across any type or scheme was the MLP NNrank ordinal classifier with an accuracy of 71\%.
For the NNRank applied to the MLP and CNN, the average test accuracy of the CNN and MLP improved minimally with +2\% and +4\%, repectively, over the nominal classification.
This improvement is accounted for mainly in reduced classification errors of the 1000$^\circ$C images from 75\% to 82\% accuracy (Figure~\ref{testcm_ordi}).

\begin{figure}
     \centering
     \includegraphics[width=0.5\textwidth]{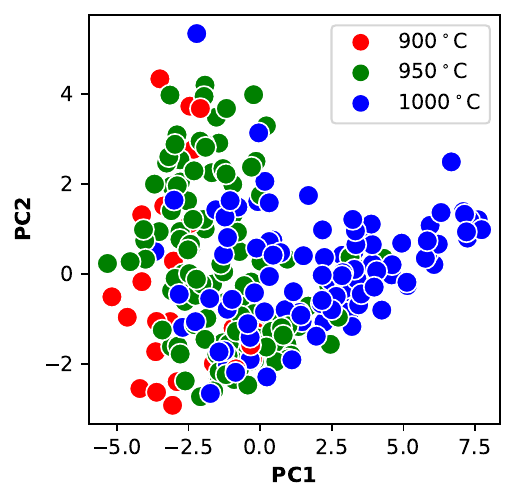}
         \caption{The first two principal components of the image features showing the temperature class distribution in the reduced dimensional representation from the principal component analysis.
         Significant overlap is observed among the different classes in the embedding space.}
         \label{pca}
\end{figure}

\begin{figure}
     \centering
     \includegraphics[width=\textwidth]{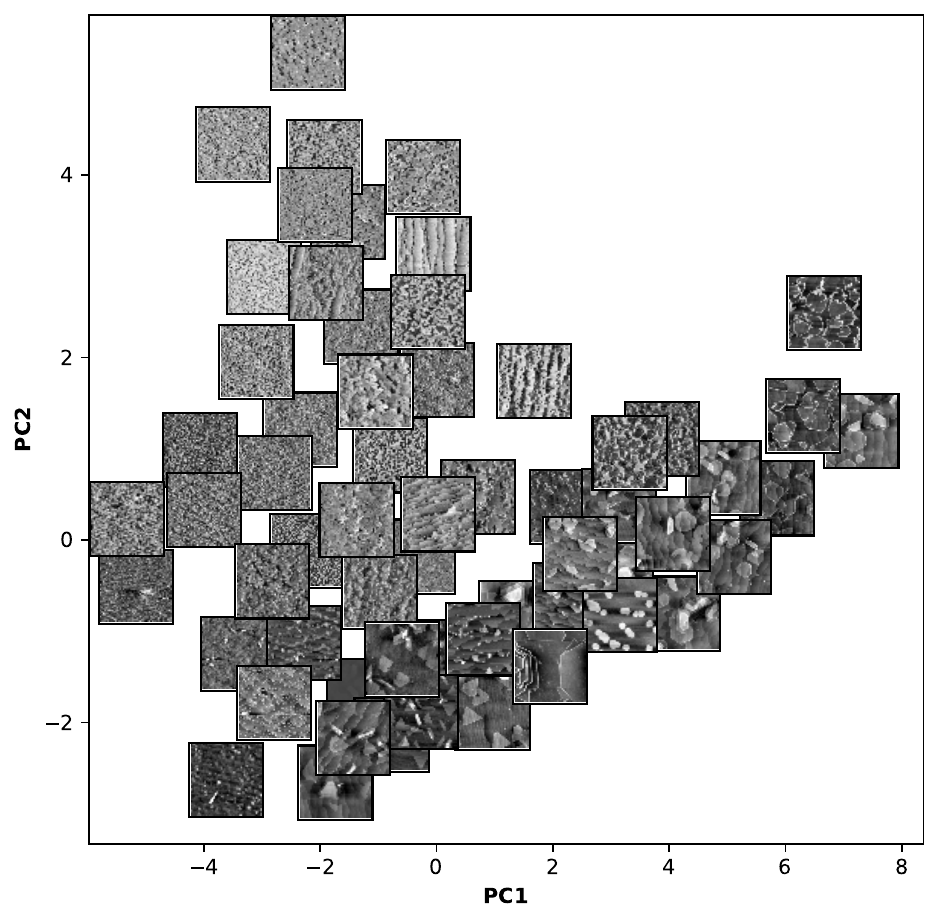}
         \caption{The first two principal components of the image features showing the sample images, in the reduced dimensional representation from the principal component analysis.
         The embedding shows that the first dimension (PC1) is associated with the domain size, while the second dimension (PC2) seems to indicate the domain density.}
         \label{pca_image}
\end{figure}

In an effort to explain the surprising trend observed in the ordinal treatment of the data, we obtained the first 2 principal components of the data using principal component analysis (PCA)\cite{pearson1901liii}.
The image classes are embedded in the 2 components shown in Figure~\ref{pca}.
The figure shows overlap of all three classes and more significantly between neighboring classes, with very poor separation visible in the first two components.
We visualize the micrographs in the PCA space in Figure~\ref{pca_image}, indicating variations in the domain size (PC1) and density (PC2).
Because these features vary significantly even within the same temperature class (e.g., see Figure~\ref{pca}), the image feature vectors likely do not show consistent trends from 900$^\circ$C to 950$^\circ$C to 1000$^\circ$C, leading to no advantage in the ordinal treatment of growth temperature.

\subsection{Model Explanations}

Beyond the capacity of the ML models to isolate the morphological features associated with the different growth temperature of the thin film \ce{MoS2} based on their AFM images, we want to understand what features of the images the models used in the classification.
Class activation maps (CAM) of the different classes are therefore obtained following the implementation by Zhou, et al.\cite{CAMzhou2016cvpr}
The feature maps of the last convolutional layer are summed and then normalized by dividing by the maximum value to obtain a heatmap with the same dimensions as the layer.
The bright yellow spot on the class activation maps represent the region with the highest activation which the model used for the classification. 

\begin{figure}
     \centering
     \includegraphics[width=0.55\textwidth]{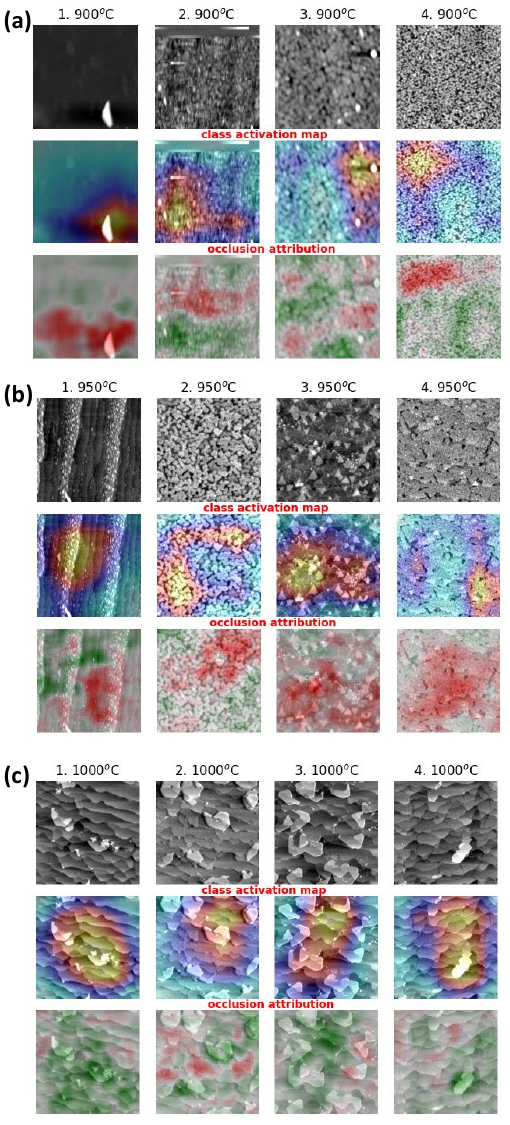}
         \caption{Class activation maps (CAM) and occlusion attribution showing different regions of the images the model used for the classification.
         (a), (b), and (c) are for samples of images grown at 900$^\circ$C, 950$^\circ$C and 1000$^\circ$C, respectively. (i), (ii), and (iii) are the original AFM images, CAM, and AFM images overlaid with CAM, respectively.}
         \label{cam}
\end{figure}

Additionally, we obtained the occlusion attribution; the probability of a class of image as a function of an occluder object,\cite{DBLP:journals/corr/ZeilerF13} using the implementation in Captum library.\cite{DBLP:journals/corr/abs-2009-07896}
To achieve this, we iteratively set a patch of the image to be zero-pixel values and then obtain the probability of the class.
Stride size of $5 \times 5$ and the patch size of $15 \times 15$ were used.
The probability is visualized as a 2D heat map.
Both positive and negative attributions, indicating that the presence and absence of the area, respectively, increases the prediction scores are shown on the heat map.
The occlusion attribution is applied to four sample images, for each class, correctly predicted by the CNN model.
Green regions on the image have positive attributions while red regions have negative attribution.

The CAM and occlusion attribution in Figure~\ref{cam} show substantial agreement in identifying the activation region, with the latter giving more specific spatial attribution.
The activation features are easier to perceive in images with bigger domain sizes, especially those grown at higher temperature.
For some of the images from samples grown at higher temperature and which show clearly defined domains, some domain boundaries are highlighted, indicating the model's reliance on the boundaries in identifying such images.
Also, regions with clean multi-steps crystals are shown to be important for the model in the classification (Figure~\ref{cam}c), while the messy crystals post adverse effect to class attribution, as shown in the occlusion attribution. 

From the experimental observations, the samples grown at higher temperatures are expected to exhibit greater domain sizes.\cite{1eichfeld2015highly, 5zhang2016influence}
However, in the data used in training our models, there is significant variation in the quality of the samples, such that most images grown at higher temperature do not necessarily have greater domain sizes (Figures~\ref{sample_AFM} and \ref{test_data}).
Additionally, if the model depends on domain size in identifying the images, it will be difficult to visually identify such features in images with less defined domains, and the only difference among the classes would only be the magnitude of the same feature.
This is unlike the natural images where activation of different classes are typically associated with unique features of the classes that can be visually identified.\cite{CAM8004993, CAM9206626, CAM9462463}
The models have therefore shown to be capable of identifying image features that humans could potentially miss.

\section{Conclusion}

This study focuses on the development of ML models for the classification of AFM images of thin film \ce{MoS2} based on the growth temperatures of their samples.
Many different strategies were explored for generating feature vectors, including using different pretraining image domains, extracting features from different depths in a pretrained ResNet, and end-to-end fine-tuning. A novel approach to transfer learning where the convolutional filters of the pretrained model were first fine-tuned before using them to extract features was also introduced. Our scheme yielded better results than the traditional approaches. Different augmentation strategies from the literature were evaluated to determine their effect on overall model performance.
Beyond these pretraining schemes, nine different ML algorithms were evaluated to determine the most suitable approach for identifying morphological features associated with different growth temperatures.

The study also examined the impact of considering the ordinality of the classes on the accuracy of the models in identifying AFM images grown at different temperatures.
We found that accounting for ordinality (i.e., by switching from classification to regression loss functions) improved the accuracy of some algorithms while decreasing performance for others.
For instance, the best model overall was obtained using an NNrank ordinal classifier, but some nominal classifier were nearly as accurate.
Furthermore, some algorithms had equivalent accuracy regardless of whether the data was treated as nominal classes or ordinal.
Thus, there seems to be no clear advantage to using least-squares regression here, despite the data appearing in the form of continuous, ordered growth temperatures, which is a counterintuitive result.

To address class imbalance, weighted random sampling and oversampling techniques were employed, and robust ML models that generalize well to out-of-sample data were developed using model ensembles.
The best-performing algorithms, MLP and end-to-end CNN, achieved classification accuracy of about 70\% on held-out test data.
The high accuracy obtained demonstrates the effectiveness of ML in accurately identifying thin films grown at different temperatures, despite the limitations of other inconsistent growth parameters and imbalances in the training data.

This study also sought to understand the features utilized by the ML models for classification by obtaining class activation maps and occlusion attribution.
These strategies revealed that images from samples grown at higher temperatures, exhibiting well-defined domains, had the highest activation at the domain boundaries, aligning with experimental observations.
Moreover, the models demonstrated the capability to identify latent features that humans could potentially miss, accurately classifying images with varying domain sizes that would be challenging for human experts.
Future work may explore the relationship between these image features and additional attributes of the samples; the robustness of these features across growth chambers, characterization instruments, and even repeatability over time may be interesting ways to utilize the quantitative capability of deep learning to unlock new insights into challenging materials synthesis problems.

\section*{Acknowledgments}

This study is based upon research conducted at The Pennsylvania State University Two-Dimensional Crystal Consortium – Materials Innovation Platform (2DCC-MIP) which is supported by NSF cooperative agreement DMR-2039351.

\section*{Data Availability}
The raw data required to reproduce these findings are available to download from Ref.\cite{Moses_Reinhart} The processed data required to reproduce these findings are available to download from Ref.\cite{moses_isaiah_a_2023_8432222} The codes used for this work can be accessed at https://zenodo.org/records/10534837
\clearpage
\bibliography{main}
\end{document}